\documentclass[twocolumn]{enoc08}
\usepackage{graphicx,array}
\usepackage{dcolumn}
\usepackage{bm}
\usepackage{amsmath,amssymb,blackboard}
\newtheorem{example}{Example}[section]

\newtheorem{theorem}{Theorem}[section]
\newtheorem{proposition}{Proposition}[section]
\newtheorem{lemma}{Lemma}[section]

\def\ip#1#2{\langle #1 | #2 \rangle}

\def\diag{\operatorname{diag}}
\def\dim{\operatorname{dim}}

\def\Tr{\operatorname{Tr}}
\def\Ad{\operatorname{Ad}}

\def\B{\mathcal{B}}
\def\C{\mathcal{C}}
\def\H{\mathcal{H}}
\def\M{\mathcal{M}}
\def\O{\mathcal{O}}
\def\T{\mathcal{T}}

\def\su{\mathfrak{su}}

\begin{document}

\title{Analysis of Lyapunov Control for Hamiltonian Quantum Systems}
\author{Xiaoting Wang$^1$ and Sonia Schirmer$^{1,2}$
\affiliation{$^1$Dept of Applied Maths \& Theoretical Physics,
             University of Cambridge, Cambridge, CB3 0WA, UK \\
             $^2$Department of Maths and Statistics, 
             University of Kuopio, PO Box 1627, 70211 Kuopio, Finland\\
             xw233@cam.ac.uk, sgs29@cam.ac.uk}}
\date{\today}
\maketitle

\begin{abstract}
We present detailed analysis of the convergence properties and
effectiveness of Lyapunov control design for bilinear Hamiltonian 
quantum systems based on the application of LaSalle's invariance 
principle and stability analysis from dynamical systems and control 
theory.  For a certain class of Hamiltonians, strong convergence
results can be obtained for both pure and mixed state systems.
The control Hamiltonians for realistic physical systems, however,
generally do not fall in this class.  It is shown that the 
effectiveness of Lyapunov control design in this case is 
significantly diminished.
\end{abstract}


\section{Introduction}

Control of quantum phenomena is becoming increasingly important in many
divergent areas of research including quantum computation, quantum
chemistry, nano-scale materials, and Bose-Einstein condensates.
Accordingly, quantum control theory has developed greatly in both depth
and breath in recent years, and many results about controllability and
control methods have been obtained.  Broadly speaking, quantum control
approaches fall into two categories: open-loop Hamiltonian (and
sometimes reservoir) engineering using a variety of techniques including
optimal control~\cite{Shi1988,Maday2003,schirmer1} and geometric
designs~\cite{jurdjevic97,jurdjevic1,lowenthal,d'alessandro,schirmer},
and closed-loop quantum state reduction and stabilization using feedback
from weak measurements~\cite{wiseman1,wiseman2}.

Lyapunov functions have played a significant role in control design.
Originally used in feedback control to analyze the stability of the 
controlled system, they have formed the basis for new control designs,
and several recent papers have discussed the application of Lyapunov
control designs to quantum systems~\cite{Vettori,Ferrante,Grivopoulos,%
Mirrahimi2004a,Mirrahimi2004b,Mirrahimi2005,altafini1,altafini2}.
Although the basic mathematical formalism is well established,
using either the Schrodinger equation for pure state wavefunctions%
~\cite{Vettori,Ferrante,Grivopoulos}, or the Liouville-von Neumann 
equation~\cite{von-Neumann} for density operators~\cite{altafini1,%
altafini2}, many questions remain.  For example, some sufficient 
conditions for the method to be effective, i.e., guarantee convergence
of the system state converges to the target state, have been
obtained~\cite{Mirrahimi2004a,Mirrahimi2004b,Mirrahimi2005,altafini1,altafini2},
but are they also necessary? What are explicit requirements on the
Hamiltonian and the target state such that the control is effective?

Moreover, there are outstanding technical issues.  The invariance 
principle~\cite{lasalle} should be applied to autonomous systems, 
and its application for time-dependent target states needs careful
justification.  Moreover, the trajectory of time-dependent target 
state under free evolution is generally not periodic, as previously 
asserted in the literature~\cite{altafini2}.  We address these issues
and present a detailed analysis of the relationship between the 
effectiveness of Lyapunov control design and the parameters in the 
control problem, the Hamiltonian and the target state.  In section 
II, we establish the mathematical model for the quantum system 
controlled by Lyapunov method, and apply LaSalle invariance principle 
to analyze the convergence condition. In section III and IV, we will 
use the results in section II to discuss the Lyapunov control of 
pseudo-pure state and generic state, under an ideal condition of 
the Hamiltonian; in section V, we shall relax this condition and 
see how the effectiveness of Lyapunov control will change.

\section{Control System and Invariance Principle}

\subsection{Controlled dynamics and Lyapunov function}

A controlled quantum system can be modelled in different ways, either as
a closed system evolving unitarily under certain Hamiltonian, or as an
open system interacting with a heat bath.  In this paper, we restrict our
discussion to an $n$-level bilinear Hamiltonian dynamical system
satisfying the Liouville-von Neumann equation: (assuming $\hbar=1$):
\begin{equation} 
\label{eqn:1}
  \dot{\rho}(t) =-i [ H_0+f(t)H_1, \rho(t) ],
\end{equation}
where $\rho$ is a positive trace-one operator, representing the
system state, $H_0$ is the free evolution Hamiltonian and $H_1$ is
the controlled Hamiltonian, both of which are constant.  In the 
special case when the system is in a pure state $|\psi \rangle$, we 
have $\rho=|\psi \rangle \langle \psi|$, and the dynamical
system can be represented as:
\begin{equation}
  \dot{\psi}(t) =-i \left(H_0+f(t)H_1\right) \psi(t)
\end{equation}
although we will use the density operator formulation throughout this
paper.  The general control task we consider can be formulated as, given 
a target state $\rho_d$, we wish to apply a certain control field $f(t)$ 
to the system that modifies its dynamics such that $\rho(t)\to\rho_d$ 
as $t\to+\infty$.  Since the free Hamiltonian $H_0$ can generally not
be turned off, it is natural to assume $\rho_d$ to be time-dependent, 
satisfying:
\begin{equation}
\label{eqn:3}
  \dot{\rho}_d(t) = -i [ H_0, \rho_d(t) ].
\end{equation}
Since the evolution of both $\rho(t)$ and $\rho_d(t)$ is unitary in our 
case, we require $\rho(0)$ to be unitarily equivalent to $\rho_d(0)$. 
Hence, the state space $\M$ is the set of all density operators $\rho$ 
such that $\rho$ and $\rho_d(0)$ are unitarily equivalent.  This is a 
compact manifold, called a flag manifold, whose dimension depends on 
the number of distinct eigenvalues of $\rho_d$.  We say $\rho_d$ is 
pseudo-pure if its spectrum has only two distinct values, one occurring 
with multiplicity one and the other with multiplicity $n-1$, and
$\rho_d$ is generic if it has $n$ non-degenerate eigenvalues.  We have
$2n-2\le\dim(\M)\le n^2-n$ with $\dim(\M)=n^2-n$ for generic $\rho_d$
and $\dim(\M)=2n-2$ for pure or pseudo-pure states.

Define a function $V$ on $\M\times\M$:
\begin{align}
V(\rho,\rho_d)&= \frac{1}{2}\Tr((\rho-\rho_d)^2)\nonumber\\
              &= \Tr(\rho_d^2)-\Tr(\rho_d\rho). \label{eqn:4}
\end{align}
We have $V\ge 0$ with equality only if $\rho = \rho_d$. Taking
derivative of $V$ along any solution $(\rho(t),\rho_d(t))$, and
substituting (\ref{eqn:1}) and (\ref{eqn:3}), we derive:
\begin{align*}
\dot V =&-\Tr(\dot \rho_d \rho)-\Tr(\rho_d \dot \rho)\\
       =&-\Tr([-iH_0,\rho_d]\rho)-\Tr(\rho_d[-iH_0,\rho])\\
        &-f(t)\Tr(\rho_d[-iH_1,\rho])\\
       =&-f(t)\Tr(\rho_d[-iH_1,\rho]).
\end{align*}
If we choose $f(t)=\kappa\Tr(\rho_d[-iH_1,\rho])$, $\kappa >0$, then
$\dot V(\rho(t),\rho_d(t)) \le 0$. Hence, $V$ is a Lyapunov function
for the following autonomous dynamical system with respect to
$(\rho(t),\rho_d(t))$:
\begin{equation}
\label{eqn:auto}
\begin{split}
  \dot{\rho}(t)   &=-i [ H_0+f(\rho,\rho_d)H_1, \rho(t) ]\\
  \dot{\rho}_d(t) &=-i [ H_0, \rho_d(t) ]\\
  f(\rho,\rho_d)  &=\kappa \Tr([-iH_1,\rho]\rho_d)
\end{split}
\end{equation}

\subsection{LaSalle invariance principle and invariant set}

To complete the control task, we require $\rho(t)\to \rho_d(t)$ as
$t\to +\infty$, which is equivalent to $V(\rho(t),\rho_d(t))\to 0$.
A key result for the convergence analysis is LaSalle's invariance
principle~\cite{lasalle}:

\begin{theorem}
\label{thm:1}
Let $V(x)$ be a Lyapunov function on the phase space $\Omega=\{x\}$
of an autonomous dynamical system $\dot x=f(x)$, satisfying $V(x)>0$ 
for all $x \neq x_0$ and $\dot{V}(x) \le 0$.  Let $\O(\bar x(t))$ be 
the orbit of $\bar x(t)$ in $\Omega$.  Then the invariant set 
$E=\{\O(x(t))|\dot{V}(\bar x(t))=0\}$, contains the positive limiting 
sets of all bounded solutions, i.e., any bounded solution converges to 
$E$ as $t \to + \infty$.
\end{theorem}
For our quantum dynamical system (\ref{eqn:auto}), since the state
space $\mathcal{M}$ is compact, any solution $(\rho(t),\rho_d(t))$
is bounded. Applying LaSalle Invariance Principle we
obtain~\cite{xiaoting}:
\begin{theorem}
The state $(\rho(t),\rho_d(t))$ of the autonomous dynamical system%
~(\ref{eqn:auto}) converges to the invariant 
set $E=\{(\rho_1,\rho_2)\in \M\times\M|\dot{V}(\rho(t),\rho_d(t))=0,
(\rho(0),\rho_d(0))=(\rho_1,\rho_2)\}$.
\end{theorem}
Therefore, the next step is to determine the invariant set $E$, for
the dynamical system~(\ref{eqn:auto}).  Notice that in LaSalle
invariance principle, $E$ contains the positive limiting points of
all bounded solutions for any $\rho_d(t)$. Hence, in the following,
we always restrict the calculation of $E$ to the points
$(\rho_1,\rho_2)$ such that $\rho_2$ is the positive limiting point
of the given $\rho_d(t)$. For our dynamical system, the invariant
set $E=\{\dot V(\rho(t),\rho_d(t))=0\}$ is equivalent to $f(t)=0$,
for any $t$:
\begin{align*}
0 &= f= \Tr([-iH_1,\rho]\rho_d)\\
0 &= \dot f= \Tr([-iH_1,\rho]\dot \rho_d)+\Tr([-iH_1,\dot \rho]\rho_d)\\
  &= -\Tr([[-iH_0,-iH_1],\rho]\rho_d)\\
  &\cdots \\
0 &= \frac{d^{\ell}f}{dt^\ell}=(-1)^n \Tr([{Ad}^\ell_{-iH_0}(-iH_1),\rho]\rho_d),
\end{align*}
where ${Ad}^\ell_{-iH_0}(-iH_1)$ represents $\ell$-fold commutator
adjoint action of $-iH_0$ on $-iH_1$.  Hence, 
$\Tr([A,B]C)=-\Tr([C,B]A)=-\Tr([A,C]B)$ gives a necessary condition
for the invariant set $E$:
\begin{equation}
\label{eq:trace-cond1}
 \Tr([\rho,\rho_d]\Ad^m_{-iH_0}(-iH_1))=0,
\end{equation}
where $\Ad^0_{-iH_0}(-iH_1)=-iH_1$ and $m$ is any non-negative
integer. Hence the invariant $E$ depends on both the Hamiltonian and
the target state. Without loss of generality, we assume $H_0$ and
$H_1$ to be trace-zero.  Since $H_0$ is hermitian and therefore 
diagonalizable, we may assume $H_0=\diag(a_1,\ldots,a_n)$, with 
diagonal elements arranged in decreasing order, where the diagonal 
elements physically represent the energy levels of the system.  We 
shall assume $H_1$ to be off-diagonal in this basis with off-diagonal 
elements $b_{k\ell}$ representing the couplings between the energy 
levels $k$ and $\ell$.  Also let $\omega_{k\ell}=a_\ell-a_k$ be the 
transition frequency between the energy levels $k$ and $\ell$.  With 
these assumptions, we can prove the following useful theorem~\cite{xiaoting}:
\begin{theorem}
\label{thm:invariant}
If (1) $H_0$ strongly regular, i.e.,
    $\omega_{k\ell}\neq \omega_{pq}$ unless $(k,\ell)=(p,q)$, and
(2) $H_1$ fully connected, i.e., $b_{k\ell}\neq 0$ except for 
    $k=\ell$, then 
$(\rho_1,\rho_2)$ belongs to the invariant set $E$ if and only if
$[\rho_1,\rho_2]$ is diagonal.
\end{theorem}
We note that these conditions on the Hamiltonian are very strong 
and rarely satisfied for real physical systems.  However, this is 
the ideal case for Lyapunov control design, and it is useful to 
begin by analyzing the effectiveness of the method in this ideal 
case before relaxing the requirements.  

\subsection{Real representation for quantum systems}

In order to apply stability analysis to our complex quantum
dynamical system, we require a real representation for both the
Hamiltonian and the density operator. Let $\B_\RR(\H)$ be the real
vector space of all $n\times n$ Hermitian matrices on the Hilbert 
space $\H$.  For any $H_1,H_2\in \B_\RR(\H)$, we can define an inner 
product $\ip{H_1}{H_2}=\Tr(H_1H_2)$, and an associated orthonormal 
basis $\{\lambda_k,\lambda_{k\ell},\bar{\lambda}_{k\ell}\}$, where
\begin{subequations} 
\label{eq:lambda}
\begin{align} 
 \lambda_0   
&= \textstyle\frac{1}{\sqrt{n}} (\hat{e}_{11}+\hat{e}_{22}+\cdots+\hat{e}_{n,n})\\
 \lambda_k   
&= \textstyle\frac{1}{\sqrt{k(k+1)}}(\hat{e}_{11}+\cdots+\hat{e}_{kk}-k\hat{e}_{k+1,k+1}) \\
 \lambda_{k\ell}      
&= \textstyle\frac{1}{\sqrt{2}}(\hat{e}_{k\ell}+\hat{e}_{\ell k})\\
 \bar{\lambda}_{k\ell}
&= \textstyle\frac{i}{\sqrt{2}}(-\hat{e}_{k\ell}+\hat{e}_{\ell k}),
\end{align}
\end{subequations}
$\hat{e}_{k\ell}$ being the elementary matrix with $1$ in the
$(k,\ell)$ position and $0$ elsewhere, and $1\le k < \ell\le n$. 
In this basis, any hermitian matrix $H$ can be represented as an
$n^2$-dimensional real vector.  For density operators $\rho$ with 
$\Tr(\rho)=1$ the coefficient of $\lambda_0$ is constant and thus 
can be dropped.  Let $\vec{s}(t)$ and $\vec{s}_d(t)$ be the vectors 
in $\RR^{n^2-1}$ representing $\rho(t)$ and $\rho_d(t)$.   
The adjoint action $\Ad_{iH}(\rho)=[iH,\rho]$ in this basis is 
given by an anti-symmetric matrix $A$ acting on $\vec{s}(t)$. 
Therefore, the quantum dynamical system~(\ref{eqn:auto}) can be 
equivalently represented as:
\begin{equation}
\label{eqn:real}
\begin{split}
\dot{\vec{s}}(t)   &= (A_0+A_1) \vec{s}(t)\\
\dot{\vec{s}}_d(t) &= A_0 \vec{s}_d(t)\\
f(t)               &= \vec{s}_d^T A_1 \vec{s}
\end{split}
\end{equation}
where $A_0=\Ad_{-iH_0}$ and $A_1=\Ad_{-iH_1}$.  This real representation
is generally known as the Bloch vector form especially for $n=2$.

\section{Convergence Analysis for Ideal Systems}

In this section we consider ideal systems, i.e., systems with $H_0$
strongly regular and $H_1$ fully connected.  If the system and hence the
Hamiltonian are fixed, the invariant set $E$ depends on the target
state $\rho_d$ only.  In general, the convergence analysis depends on
the spectrum of the target state $\rho_d$, in particular the
multiplicities of its eigenvalues.  As we cannot give a complete
discussion of all possible cases here, we will focus on two cases of
crucial importance, (a) when $\rho_d$ is a generic mixed state with $n$
non-degenerate eigenvalues, and (b) when $\rho_d$ is a pure or
pseudo-pure state with only two eigenvalues with multiplicities $1$ and
$n-1$, respectively, starting with the two-level case.

\subsection{Pseudo-pure states}

We start with the special case of a two-level system, for which the 
embedding of density operators into $\RR^{n^2-1}$ gives rise to a 
\emph{homeomorphism} between density operators and points inside a 
closed ball in $\RR^3$, with pure states forming the surface of the 
ball, and the completely mixed state $(0,0,0)$ its centre.  In this
case there is no distinction between pseudo-pure and generic states, 
all states except the completely mixed state 
$\rho_0=\diag(\frac{1}{2},\frac{1}{2})$ 
being both generic and pseudo-pure.  Furthermore, the requirements of 
strong regularity of $H_0$ and full connectedness of $H_1$ are always
satisfied for a two-level systems, except for the trivial cases of 
systems with a single degenerate state or no coupling to the control 
field.  Excluding these trivial cases, Theorem~\ref{thm:invariant} 
implies that all $(\rho_1,\rho_2)\in E$ satisfy $[\rho_1,\rho_2]$ 
diagonal, and one can show that there are three types of points in 
the invariant set~\cite{xiaoting}
\begin{itemize}
\item[(a)] $\Tr(\rho_1\rho_2)=1$, i.e., $\rho_1=\rho_2$;
\item[(b)] $\Tr(\rho_1\rho_2)=0$;
\item[(c)] $\Tr[\lambda_1\rho_1] = \Tr[\lambda_1\rho_2]=0$.
\end{itemize}
Another special feature of the $n=2$ case is that for any choice
of $H_0$, the trajectory of $\rho_d(t)$ forms a periodic orbit
$\O(\rho_d(0))$, which is a compact set, so any positive limiting 
point $(\rho_1,\rho_2)$ of $(\rho(t),\rho_d(t))$ must satisfy 
$\rho_2\in\O(\rho_d(0))$.  Therefore, if $\rho_d(0)$ has nonzero 
$\lambda_1$ component, i.e., $\Tr(\rho_d(0)\lambda_1)\neq 0$ then 
$E$ can only contain the cases (a) and (b), corresponding to the 
values of Lyapunov function $V=V_{\rm max}$ and $V=0$.  Hence, 
for any $\rho(t)$ with $\Tr(\rho(0)\rho_d(0)) \neq 0$, we have 
$V(\rho(0),\rho_d(0))<V_{\rm max}$ and LaSalle's invariance principle
guarantees that $V(\rho(t),\rho_d(t))\to 0$ and $\rho(t)\to\rho_d(t)$ 
as $t\to+\infty$.  Lyapunov control in this case is an effective
strategy.

If $\rho_d(0)$ has zero $\lambda_1$ component, however, then the
invariant set $E$ contains all points $(\rho_1\rho_2)$ satisfying
(c), the value of the Lyapunov function $V$ on $E$ spans the entire 
interval $[0,V_{\rm max}]$, and we cannot conclude that $\rho(t) 
\to\rho_d(t)$.  Indeed simulations suggest that $V$ can tend to any
value in $[0,V_{\rm max}]$ in this case.  We can still conclude that 
$\rho(t)$ converges to the set $\O(\rho_d(t))$ corresponding to the 
orbit of $\rho_d(t)$ but this is a substantially weaker notion of 
convergence as there are infinitely many distinct states whose orbits
under free evolution coincide.  

If we take the Bloch vector to be $\vec{s}=(x,y,z)$ with 
$x=\Tr(\rho\lambda_{12})$, $y=\Tr(\rho\bar\lambda_{12})$ and
$z=\Tr(\rho\lambda_1)$, as usual for $n=2$, then case (a) corresponds 
to $\vec{s}_1=\vec{s}_2$, case (b) corresponds to $\vec{s}_1$ being
antipodal to $\vec{s}_2$, $\vec{s}_1=-\vec{s}_2$, and (c) corresponds 
to the target state lying on the equator of the sphere.  If 
$\vec{s}_d(0)$ is not on the equator, all solutions $\vec{s}(t)$ with 
$\vec{s}(0)\ne -\vec{s}_d(0)$ converge to $\vec{s}_d(t)$.  If 
$\vec{s}_d(0)$ is on the equator, any solution $\vec{s}(t)$ will 
converge to the equator but $\vec{s}(t)\not\to\vec{s}_d(t)$ in general.  
The picture is the same for all equivalence classes of states, except 
the completely mixed state, the only difference being that pure states 
lie on the surface of the ball, while pseudo-pure states with the same 
spectrum lie on concentric spherical shells of in the interior. 

For $n>2$ pseudo-pure states are exceptional or non-generic and the
mapping from density operators into $\RR^{n^2-1}$ provided by the
Bloch vector is only an embedding, not a homeomorphism.  Furthermore,
the conditions on the Hamiltonian of strong regularity and complete
connectedness are less trivial in this case as there are many systems
that are connected and regular and controllable, but not strongly
regular or fully connected.  However, assuming such ideal Hamiltonians 
we can still prove~\cite{xiaoting}:

\begin{theorem}
\label{thm:conv:ideal-pseudo-pure}
Given a pseudo-pure state target state $\rho_d(t)$ with spectrum $\{w,u\}$ 
and ideal Hamiltonians as defined above, Lyapunov control is effective, i.e., 
any solution $\rho(t)$ with $V(\rho(0),\rho_d(0))<V_{\rm max}$ will converge 
to $\rho_d(t)$ as $t\to+\infty$, \emph{except} when $\rho_d$ has a single 
pair of non-zero off-diagonal entries of the form 
$r_{k\ell}(t)=\frac{1}{2}(w-u)e^{i\alpha}$ and 
$r_{kk}=r_{\ell\ell}=\frac{1}{2}(w+u)$.  In the latter case any solution 
$\rho(t)$ will converge to the orbit of $\rho_d(t)$ but in general 
$\rho(t)\not\to\rho_d(t)$ as $t\to +\infty$ and $V(\rho,\rho_d)$ can take 
any limiting value between $0$ and $V_{\rm max}$.
\end{theorem}

This theorem essentially asserts that if $H_0$ is strongly regular, $H_1$
is fully connected and $\rho_d$ is a pseudo-pure state whose dynamics is
not confined to a periodic orbit in a two-dimensional subspace, then any
solution $\rho(t)$ with $\Tr(\rho(0)\rho_d(0))\ne 0$ will converge to
$\rho_d(t)$ as $t\to+\infty$.  Thus for $n>2$ the special case where
$\rho(t)$ converges to the orbit $\O(\rho_d(t))$ but not $\rho_d(t)$,
corresponding to case (c) above, is precluded, except when the target
state is such that its orbit is a periodic orbit (circle) in a in a 2D
subspace.  Given any other pseudo-pure state $\rho_d(t)$, all initial
states $\rho(0)$ that are not part of the critical manifold of states 
for which $\Tr(\rho(0)\rho_d(0))=0$ and $V$ assumes its maximum, will 
converge to the target state $\rho_d(t)$ for $t\to\infty$.

\subsection{Generic states for $n$-level systems}

For $n>2$ pseudo-pure states are a very small subset of the state space.
Most states $\rho$ are generic with $n$ non-degenerate eigenvalues.  We
distinguish two cases here: stationary target states $\rho_d$, which are
diagonal (in the eigenbasis of $H_0$), and non-stationary target states.
When $\rho_d$ is stationary, the dynamical system (\ref{eqn:auto}) can 
be reduced to
\begin{equation}
\label{eqn:auto1}
\begin{split}
 \dot\rho(t) &= -i [ H_0+f(\rho)H_1, \rho(t) ]\\
 f(\rho)     &= \Tr([-iH_1,\rho(t)]\rho_d)
\end{split}
\end{equation}
and the invariant set (for an ideal system) reduces accordingly to 
$E=\{\rho_0|\dot V_{\rho_d}(\rho(t))=0,\rho(0)=\rho_0\}$, which can be 
shown to be equivalent to the set of all $\rho_0$ with $[\rho_0,\rho_d]$ 
diagonal.  Furthermore, since $\rho_d$ is stationary and thus diagonal,
the latter condition can further be reduced to $[\rho_0,\rho_d]=0$ by 
virtue of the following:

\begin{lemma}
\label{lemma:1}
If $A$ is diagonal with non-degenerate eigenvalues and $[A,B]$ is
diagonal, then $B$ is also diagonal and $[A,B]=0$.
\end{lemma}

The proof follows trivially from the fact that for
$A=\diag(a_1,\ldots,a_n)$ and $B=(b_{mn})$, the $(m,n)$ component of
$[A,B]$ is $b_{mn}(a_m-a_n)$.  If $[A,B]$ is diagonal and $a_m\neq a_n$
then we must have $b_{mn}=0$ for $m\ne n$.  This leads to the following~\cite{xiaoting}:

\begin{theorem}
\label{thm:generic:crit}
If $\rho_d$ is a generic stationary target state then the invariant set 
$E$ contains exactly the $n!$ critical points of the Lyapunov function 
$V(\rho)=\Tr(\rho_d^2)-\Tr(\rho\rho_d)$, i.e., the stationary states 
$\rho_d^{(k)}$, $k=1,\ldots,n!$, which commute with $\rho_d$ and have the 
same spectrum.  
\end{theorem}

As $\rho_d$ is stationary and therefore diagonal, it follows that all
$\rho_d^{(k)}$ are also diagonal, and their diagonal elements are a
permutation of those of $\rho_d$.  Since $\Tr(\rho_d^2)$ is constant for
a Hamiltonian system, the critical points of $V(\rho)$ coincide with the
critical points of $J(\rho)=\Tr(\rho_d\rho)$, which can be regarded as
the expectation value of the observable $A=\rho_d$.  We can immediately
see that $V$ assumes its global minimum when the expectation value of
$\Tr(\rho_d\rho)$ assumes its maximum, i.e., for $\rho=\rho_d$, and its
maximum when $J(\rho)$ assumes its minimum.  Assuming
$\rho_d^{(0)}=\rho_d=\diag(w_1,\ldots,w_n)$ and
$\rho_d^{(n!)}=\diag(w_{\tau(1)},\ldots,w_{\tau(n)})$, where $\tau$ is
the permutation of $\{1,\ldots,n\}$ that corresponds to a complete
inversion, i.e., $\tau(k)=n+1-k$, we have~\cite{Schirmer1998}:
\begin{equation}
  J(\rho_d^{(n!)}) \le J(\rho) \le J(\rho_d^{(1)}),
\end{equation}
i.e., $\rho_d^{(0)}$ and $\rho_d^{(n!)}$ correspond to the global extrema 
of $V$ with $V=0$ and $V=V_{\rm max}=\sum_k w_k^2$, respectively.

It is furthermore easy to show that for a given generic stationary state 
$\rho_d$ the critical points of the Lyapunov function $V(\rho)$ are 
hyperbolic.  However, since the dynamical system defined by our Lyapunov 
control is not the gradient flow of $V(\rho)$, asymptotic stability of 
these fixed points can not be derived directly from the associated index 
number of the Morse function $V$.  Nonetheless, further analysis of the 
linearization of the dynamics near the critical points shows that~\cite{xiaoting}:

\begin{theorem}
\label{thm:generic:hyperbolic}
For a generic stationary target state $\rho_d$ all the critical points 
of the dynamical system~(\ref{eqn:auto1}) are hyperbolic. $\rho_d$ is 
the only sink, all other critical points are saddles, except the global 
maximum, which is a source.  
\end{theorem}

Since the critical points of the dynamical system (\ref{eqn:auto}) for
a generic stationary state $\rho_d$ are hyperbolic and they are also 
hyperbolic critical points of the function $V(\rho)=V(\rho,\rho_d)$, 
the dimension of the stable manifold at a critical point must be the 
same as the index number of the critical point of the function $V$.
In particular, since all critical points except the global minimum and
maximum are saddle points, they are not repulsive, and therefore there 
are solutions $\rho(t)$ outside $E$ that converge to these saddles, 
resulting in the failure of the Lyapunov control method.  However, as 
the dimensions of the stable manifolds at these points are smaller than 
$\dim(\M)$, almost all solutions will still converge to the global 
minimum $\rho_d^{(1)}=\rho_d$, and thus the Lyapunov method is still 
(mostly) effective.

When the target state $\rho_d$ is not stationary, the situation is
somewhat more complicated as the invariant set $E$ may contain points
with nonzero diagonal commutators.

\begin{example} Consider $\rho_1$ and $\rho_d(0)=\rho_2$ with
\begin{equation*}
\rho_1= \begin{pmatrix}
 \frac{1}{12} & -\frac{1}{12} & -\frac{1}{12}\\
-\frac{1}{12} & \frac{11}{24} & \frac{1}{8} \\
-\frac{1}{12} & \frac{1}{8}   & \frac{11}{24}
\end{pmatrix}, \; 
\rho_2 =
\begin{pmatrix}
\frac{1}{3} & \frac{-i}{12} & \frac{i}{12}  \\
\frac{i}{12} & \frac{1}{3} & -\frac{i}{4} \\
\frac{-i}{12} & \frac{i}{4}   & \frac{1}{3}
\end{pmatrix}.
\end{equation*}
$\rho_1$ and $\rho_2$ are isospectral and we have
\begin{equation*}
[\rho_1,\rho_2]= \begin{pmatrix}
0 & 0 & 0  \\
0 & \frac{11}{144}i & 0 \\
0 & 0 & -\frac{11}{144}i
\end{pmatrix}
\end{equation*}
i.e., $(\rho_1,\rho_2)\in E$ but $[\rho_1,\rho_2]\neq 0$.
\end{example}
Simulations suggest that $\rho(t)$ does not converge to $\rho_d(t)$ or 
$\O(\rho_d(t))$ in this case and thus Lyapunov control fails.  It is 
difficult to give a rigorous proof of this observation, however, as we 
lack a constructive method to ascertain asymptotic stability near a 
non-stationary solution.  In the special case where $\rho_d(t)$ is 
periodic there are tools such as Poincar\'e maps but it is difficult 
to write down an explicit form of the Poincar\'e map for general 
periodic orbits~\cite{Perko}.  Moreover, as observed earlier,
for $n>2$ the orbits of non-stationary target states $\rho_d(t)$ under
$H_0$ are periodic only in some exceptional cases.  

Fortunately, however, $E=\{[\rho_1,\rho_2]=0\}$ still holds for a very
large class of generic target states $\rho_d(t)$, and in these cases
Lyapunov control tends to be effective.  Setting
$[\rho_1,\rho_2]=-\Ad_{\rho_2}(\rho_1)$, where $\Ad_{\rho_2}$ is a
linear map from the Hermitian or anti-Hermitian matrices into $\su(n)$,
let $A(\vec{s}_2)$ be the real $(n^2-1)\times (n^2-1)$ matrix
corresponding to the Stokes representation of $\Ad_{\rho_2}$.  Recall
$\su(n)=\T\oplus\C$ and $\RR^{n^2-1}=S_\T\oplus S_\C$, where $S_\C$ and
$S_\T$ are the real subspaces corresponding to the Cartan and non-Cartan
subspaces, $\C$ and $\T$, respectively.  Let $\tilde{A}(\vec{s}_2)$ be
the first $n^2-n$ rows of $A(\vec{s}_2)$ (whose image is $S_\T$).  
Then we can show~\cite{xiaoting}:

\begin{theorem}
The invariant set $E$ for a generic $\rho_d(t)$ contains points with 
nonzero commutator only if either $\rho_d$ has some equal diagonal 
elements or $\det(\tilde{A}_1)=0$.  Therefore, the set of $\rho_d(0)$ 
such that $E$ contains points with nonzero commutator has measure zero 
with respect to the state space $\M$.
\end{theorem}

Hence, if we choose a generic target state $\rho_d(0)$ randomly, with 
probability one, it will be such that $E=\{[\rho_1,\rho_2]=0\}$.
Choosing an orthonormal basis such that $\rho_2$ is diagonal, it thus
follows that $\rho_1$ must be diagonal in this basis, and its diagonal 
elements a permutation of the eigenvalues of $\rho_2$.  Thus for a 
given target state $\rho_2=\rho_d$, there are again $n!$ critical points 
$(\rho_d^{(k)}(t),\rho_d(t))$.  Moreover, for any $(\rho_1,\rho_2)\in E$ 
with $\rho_2=\rho_d$ and $\rho_1=\rho_2^{(k)}$ for some $k$, there 
exists a subsequence $\{t_n\}$ such that 
$(\rho(t_n),\rho_d(t_n)) \to (\rho_1,\rho_2)$.  In particular, 
$\rho(t_n)\to\rho_1$, $\rho_d(t_n)\to \rho_2$, and hence, 
$\rho_d^{(k)}(t_n)\to \rho_2^{(k)}=\rho_1$.  Therefore, we have
$\rho(t_n)\to \rho_d^{(k)}(t_n)$.
If $(\rho_1,\rho_2)\in E$ is a different positive limiting point of
$(\rho(t),\rho_d(t))$, since $V(\rho(t),\rho_d(t))$ decreases along 
the trajectory and the $n!$ critical points are isolated, we must 
have $\rho_1=\rho_2^{(k)}$ for the same $k$ as for $(\rho_1,\rho_2)$. 
Therefore, $\rho(t_n)\to \rho_d^{(k)}(t_n)$ holds for any positive 
limiting point $(\rho_1,\rho_2)$ and the corresponding subsequence 
$\{t_n\}$.  Hence $\rho(t)\to \rho_d^{(k)}(t)$ as $t\to+\infty$, and 
some further analysis shows~\cite{xiaoting}:

\begin{theorem}
\label{thm:generic:conv1}
If $\rho_d(t)$ is a generic state with invariant set
$E=\{[\rho_1,\rho_2]=0\}$ then any solution $\rho(t)$ converges to 
one of the $n!$ critical points $\rho_d^{(k)}(t)$, and all
solutions except $\rho_d^{(1)}(t)=\rho_d(t)$, which is stable, are
unstable.
\end{theorem}

Thus we have a similar result as for generic stationary $\rho_d$. 
The difference is that for the stationary $\rho_d$, we can present 
a quantitative result about the dimensions of the stable manifolds
and hence the measure of solutions that will converge to these points.
For the non-stationary target case we can not establish the analogous
result.  However, simulations suggest that almost all solutions will
converge to $\rho_d(t)$, and we can further prove a weaker 
result~\cite{xiaoting}:

\begin{proposition}
For any of the unstable solutions, $\rho_d^{(k)}$, $k=2,\ldots,n!-1$, 
we can find solutions $\rho(t)$, with 
$V(\rho(0),\rho_d(0))>V(\rho_d^{(k)}(0),\rho_d(0))$, that still 
converge to $\rho_d(t)$.
\end{proposition}

\section{Effectiveness of Method for Non-ideal Systems}

The previous analysis relied on strong assumptions about the system, 
assuming $H_0$ strongly regular and $H_1$ fully connected. 
We shall now relax these requirements to see how the invariant set $E$ 
and the effectiveness of the Lyapunov control method change.  Without 
loss of generality, we present the analysis for a three-level system, 
noting that the generalization to $n$-level systems is straightforward. 

First suppose $H_0$ is strongly regular, as for a Morse oscillator,
for example, but some of the off-diagonal elements of $H_1$ are zero,
corresponding to transitions with zero transition probability.  This
is the case for many physical systems, where often only transitions 
between adjacent energy levels are permitted, and we rarely have 
non-vanishing transition probabilities for all possible transitions.
For concreteness, assume $n=3$ and
\begin{equation*}
H_0= \begin{pmatrix}
a_1 & 0 & 0  \\
0 & a_2 & 0  \\
0 & 0 & a_3
\end{pmatrix}, \quad
H_1= \begin{pmatrix}
0 & b_1 & 0  \\
b_1^* & 0 & b_2  \\
0 & b_2^* & 0
\end{pmatrix},
\end{equation*}
where $a_1>a_2>a_3$. In this case, we can prove~\cite{xiaoting} that
any point $(\rho_1,\rho_2)\in E$ must satisfy:
\begin{equation*}
[\rho_1,\rho_2]= \begin{pmatrix}
\alpha_{11} & 0 & e^{-i\omega_{13}t}\alpha_{13}  \\
0 & \alpha_{22} & 0  \\
e^{i\omega_{13}t}\alpha_{13}^* & 0 & \alpha_{33}
\end{pmatrix}.
\end{equation*}

Hence, for a stationary and generic target state $\rho_d$ the points 
$\rho_1$ in $E$ must have the form
\begin{equation*}
\rho_1=\begin{pmatrix}
\beta_{11} & 0 & \beta_{13}  \\
0 & \beta_{22} & 0  \\
\beta_{13}^* & 0 & \beta_{33}
\end{pmatrix}.
\end{equation*}
We can prove that near the stationary point $\rho_d$, the invariant set 
$E$ forms a centre manifold with $\rho_d$ as a centre~\cite{xiaoting}. 
Therefore, the Hartman-Grobman theorem from the centre manifold theory, 
proved by Carr~\cite{Carr}, implies that almost all solutions near 
$\rho_d(t)$ converge to the (periodic) solutions on the centre manifold 
instead of the centre $\rho_d$.  Lyapunov method in this case is 
ineffective. 

Second, we consider the case of $H_1$ fully connected but $H_0$ is not 
strongly regular.  For example, consider
\begin{equation*}
H_0=\begin{pmatrix}
a_1 & 0 & 0  \\
0 & a_2 & 0  \\
0 & 0 & a_3
\end{pmatrix}, \quad
H_1=\begin{pmatrix}
0 & b_1 & b_3  \\
b_1^* & 0 & b_2  \\
b_3^* & b_2^* & 0
\end{pmatrix}
\end{equation*}
with $\omega_{12}=\omega_{23}$, where $\omega_{mn}=a_m-a_n$.  In this
case we can also prove that the invariant set $E$ forms a centre 
manifold near $\rho_d$ with $\rho_d$ as a centre, and thus that almost 
all solutions near $\rho_d(t)$ will converge to the periodic solutions 
on the centre manifold other than the centre $\rho_d$, and hence that
the Lyapunov method is still ineffective.

In summary, when either of the stringent conditions on the Hamiltonians
are relaxed even slightly, the invariant set $E$ becomes much larger, and
in marked contrast to the ideal system case, and the Lyapunov method 
fails for almost all cases.

\section{Conclusion and remarks}

We have presented a detailed analysis of the Lyapunov control method 
for bilinear quantum control systems based on the application of the
LaSalle invariance principle.  For the case of non-stationary target
states, this required considering the dynamics on an augmented state
space on which the total system is autonomous.  Characterization and
analysis of the invariant set of this dynamical system allowed us to
establish a quite clear picture of the effectiveness of the Lyapunov 
method depending on the properties of the system and the target state. 
In particular, our analysis suggests that the method is generally only 
effective under very stringent assumptions on the Hamiltonian, and even
in this case our analysis suggests a rather more complicated picture 
than previously presented in the literature (see e.g.~\cite{altafini1,altafini2}), 
in that most of the critical points, for instance, are 
unstable but \emph{not} repulsive.  

For generic stationary target states, it can be shown explicitly that all 
of the unstable critical points except the global maximum in fact have 
attractive manifolds of positive dimension.  For target states that 
are not stationary under the action of $H_0$, there are additional
complications in that the invariant set can be larger than the set of
critical points of the Lyapunov function, although the set of target
states $\rho_d(t)$ for which this happens in the ideal system case is
of measure zero.  Thus, while these issues complicate the problem for 
systems with strongly regular free (drift) Hamiltonian $H_0$ and fully 
connected control Hamiltonian $H_1$,  Lyapunov control is still
generally effective in that for most target states $\rho_d(t)$ the 
invariant set contains only the critical points of the Lyapunov function
$V$, and the target state $\rho_d(t)$ is the only hyperbolic sink of the 
dynamical system.

The situation changes radically when either of the twin requirements of 
strong regularity of $H_0$ and full connectedness of $H_1$ are relaxed, 
even slightly.  In this case the method becomes not only less effective,
but the emergence of centre manifolds around the target state suggests 
that the method is likely to become ineffective in practice.  Since the
strict requirements above do not appear to be satisfied for most physical 
systems of interest, this suggests that the utility of this method for 
practical control field design is rather limited.  It would be desirable
to have stronger analytic results for non-stationary target states where
well-established tools for stability and convergence analysis near fixed 
points are generally no longer applicable.  Although some theoretical 
tools such as the Poincare map and Floquet's theorem~\cite{Perko} exist, 
to fully answer the question of stability and convergence of Lyapunov 
control for non-stationary (and in general non-periodic) target states, 
even for ideal bilinear Hamiltonian systems, would seem to require the
development of new tools in dynamical systems theory.

\textbf{Acknowledgements:}
XW is supported by Cambridge Overseas Trust and an Elizabeth Cherry
Major Scholarship from Hughes Hall, Cambridge.  SGS acknowledges 
funding from an \mbox{EPSRC} Advanced Research Fellowship, the 
\mbox{EPSRC QIP IRC} and Hitachi, and is currently also a Marie 
Curie Fellow under the European Union Knowledge Transfer Programme 
MTDK-CT-2004-509223.  We
thank Peter Pemberton-Ross, Tsung-Lung Tsai, Christopher Taylor,
Yaxiang Yuan, Jack Waldron, Jonny Evans, Dan Jane, Jonathan Dawes,
Lluis Masanes, Rob Spekkens, Ivan Smith for helpful and interesting
discussions.

\end{document}